\documentclass[12pt,a4paper]{article}

\usepackage[numbers]{natbib}
\usepackage{rotating}
\usepackage{graphicx}
\usepackage{amssymb}

\title{Several Examples of Application of the Simple Equations  Method (SEsM) for Obtaining Exact Solutions of Nonlinear PDEs}

\author{Zlatinka I. Dimitrova}
\date{Institute of Mechanics, Bulgarian Academy of Sciences, Acad. G. Bonchev Str., Bl. 4, 1113 Sofia, Bulgaria}

\begin{document}

\maketitle

\begin{abstract}
We apply  the Simple Equations Method (SEsM) for obtaining  exact solutions of nonlinear differential equations. We discuss several examples with goal to illustrate the results from the use of derivatives of composite functions in the algorithm of SEsM. The discussed examples contain derivatives of functions 
which are composite functions of  solutions of two  simple equations.
\end{abstract}
\section{Introduction}
The  complex systems are in the most cases  non-linear ones  \cite{a1}-
\cite{a4}. Thus, large efforts are focused on the study of the effects of this nonlinearity. Such  effects   are studied  by the
methods of  the time  series analysis and by  models based on differential or 
difference equations \cite{n1} - \cite{t10}. Usually, the model equations are nonlinear partial  differential or difference equations.  Thus, the exact and approximate analytical solutions od such equation are of great interest. 
The methodology for obtaining such solutions  is in development since several decades. 
At the beginning, researchers tried
to transform the nonlinearity of the studied equation and even to remove it 
by means of appropriate transformation. One example is the Hopf-Cole 
transformation \cite{hopf}, \cite{cole}. It  transforms the nonlinear 
Burgers  equation to the linear heat equation. Following this way, one arrived at  
the Method of Inverse Scattering 
Transform \cite{ablowitz} - \cite{gardner}  the  method of Hirota  \cite{hirota}, 
\cite{hirota1}. Another line of research connected to the use of transformations is 
\cite{v23}.
\par 
Below,  we discuss and apply the SEsM (Simple Equations Method) \cite{v20} -\cite{v22a}. This method is a result of another branch of research on the methodology.  Kudryashov and then Kudryashov and Loguinova developed   the 
Method of Simplest Equation (MSE) \cite{k05},\cite{kl08} . This method is based on determination of singularity order $n$ of the solved 
nonlinear partial differential equation and searching of
particular solution of this equation as series containing powers of solutions
of a simpler equation called simplest equation. The SEsM methodology has a long story
until its recent formulation which was given in \cite{se1} - \cite{se4}. 
SEsM was proposed by Vitanov
after many years of research which started almost 35 years ago 
\cite{mv1} - \cite{mv5}. Then, in 2009
\cite{1}, \cite{2} and in  2010, Vitanov and co-authors  used the ordinary differential  equation of Bernoulli as simplest equation \cite{v10} and   applied 
the simplest version of of SEsM called Modified Method of Simplest Equation (MMSE) to ecology and population dynamics \cite{vd10}.  MMSE used a balance equation
\cite{vdk}, \cite{v11} to determine  the kind
of the simplest equation and truncation of the series of solutions of the 
simplest equation. MMSE  is equivalent of the MSE mentioned above. 
Up to 2018 the contributions to the methodology and its application have been 
connected to the MMSE  \cite{v11a} - \cite{vdv17}. This research was based on 
single simplest equation and one balance 
equation. The construction of the  solution of the solved equation was chosen to
be as power  series containing powers of the solutions of the simplest equation.

Recently Vitanov extended the capacity of 
the methodology by inclusion of the possibility of use of more than one simplest 
equation. This version  is called   SEsM - Simple Equations Method as 
the used simple equations are more simple than the solved nonlinear 
partial differential equation but these simple equations in fact can be quite 
complicated.   We note that a variant of SEsM based on two simple equations was 
applied in \cite{vd18} and the first description of the methodology was made in 
\cite{se1} and then in \cite{se2} - \cite{se4}. For more applications of specific cases of the  methodology see \cite{n17},\cite{vnew21x}. 
\par
In this article we will show  several examples of application of SEsM.
We illustrate the use of composite function which is a
function of two simple functions. Each simple functions  can be a
function of two independent variables. The structure of the article is as follows.
 We  describe  SEsM   in Sect 2. In Sect. 3, we supply the  information needed for
the use of derivatives of composite functions in SEsM. Several examples are shown in Sect. 4 and  Sect. 5 presents some concluding remarks. 
\section{The Simple Equations Method (SEsM)}
We consider a  a system of nonlinear partial differential equations
\begin{equation}\label{se1}
{\cal E}_i [u_1(x,\dots,t), \dots, u_n(x,\dots,t) ] = 0, i=1,\dots,n.
\end{equation}
${\cal E}_i [u_1(x,\dots,t), \dots, u_n(x,\dots,t) ]$ depends on the functions $u_1(x,\dots,t), \dots, u_n(x,\dots,t) $ and some of their derivatives ($u_i$ can be a function of more than 1
spatial coordinates). Step 1 of SEsM includes  transformations
\begin{equation}\label{se2}
u_i(x, . . . , t) = T_i[F_i(x, \dots, t), G_i(x,\dots,t), \dots].
\end{equation}
Here $T_i(F_i,G_i, \dots)$ is some function of another functions $ F_i,G_i,\dots$. 
Note that $T_i$ are composite functions.
In general $ F_i(x, \dots , t)$, $G_i(x,\dots,t)$, $\dots$  can be  functions of several spatial variables
as well as of the time. The  goal of the transformations is  to transform the nonlinearity of the solved differential equations to
more treatable kind of nonlinearity. In the best case, the transformation  removes 
the nonlinearity and the solved nonlinear differential equation is reduced to a linear equation. 
\par 
The nonlinearities in the solved equations can be different kinds. For a example,
for the case of one solved equation the transformation
$T(F,G,\dots)$ can be the Painleve expansion. If the solved equation has polynomial
nonlinearities one can skip this step. 
\par 
Next, one makes  Step 2. of SEsM where  the functions $F_i(x, . . . , t)$, $G_i(x,\dots,t)$, $\dots$ are 
represented as a function of other functions $f_{i1}, . . . , f_{iN}$,
$g_{i1},\dots,g_{iM}$, $\dots$,   which are connected to solutions
of some differential equations (these equations can be partial or ordinary 
differential equations) that are more
simple than Eq.(2).  The forms of the functions $F_i(f_1,\dots,f_N)$, $G_i(x,
\dots,t)$, $\dots$ can be 
different.
At Step 3. of SEsM, we choose the functions used in $F_i, G_i, \dots$ - the functions $f_{i1},\dots,f_{iN}$, $g_{i1}, \dots, g_{iM}$ are solutions of PDEs which 
are more
simple than the solved nonlinear partial differential equation. These more simple equations usually are   
ordinary differential equations. 
In many cases the form of the simple equations is determined by a balance equations. Balance
equations may be needed in order to ensure that the system of algebraic equations from Step 4.
contains more than one term in any of the equations. This is needed for a
non-trivial solution of the solved nonlinear partial differential equation.
\par 
At Step 4. of SEsM we apply  the steps 1 - 3 to Eqs.(2) and this transforms 
the left-hand side of these equations. In the most cases 
 the result of this transformation are functions which are sum of terms where 
 each term contains some
function multiplied by a coefficient. This coefficient contains some of the 
parameters of the solved equations and
some of the parameters of the solution. In the most cases a balance procedure must be applied in order to ensure
that the above-mentioned relationships for the coefficients contain more than one term ( e.g., if the result of the
transformation is a polynomial then the balance procedure has to ensure that the coefficient of each term of the
polynomial is a relationship that contains at least two terms). This balance procedure may lead to one or more
additional relationships among the parameters of the solved equation and parameters of the solution. The last
relationships are called balance equations.
\par 
Finally at Step 4. of SEsM We can obtain a nontrivial solution of Eq. (2) if all coefficients mentioned in Step 3 are set to 0. This
condition leads to a system of nonlinear algebraic equations. Each nontrivial solution of this algebraic system leads to a solution
the studied nonlinear partial differential equation. 
\section{Composite functions in SEsM}
Let us consider the function $h(x_1,\dots,x_d)$. It is a function of $d$ independent variables
$x_1,\dots,x_d$. We assume hat $h$ is a composite function of $m$ other functions
$g_1^{(1)},\dots,g^{(m)}$
\begin{equation}\label{m1}
h(x_1,\dots,x_d) = f[g^{(1)}(x_1,\dots,x_d),\dots,g^{(m)}(x_1,\dots,x_d)].
\end{equation}
Following notations are introduced.
\begin{enumerate}
\item
$\vec{\nu}=(\nu_1,\dots,\nu_d)$ is a $d$-dimensional index containing the integer non-negative 
numbers $\nu_1,\dots,\nu_d$.
\item
$\vec{z}=(z_1,\dots,z_d)$ is a $d$-dimensional object containing the real numbers $z_1,\dots,z_d$.
\item 
$\mid \vec{\nu}\mid = \sum \limits_{i=1}^d \nu_i$ is the sum of the elements of the $d$-dimensional
index $\vec{\nu}$.
\item
$\vec{\nu}! = \prod \limits_{i=1}^d \nu_i!$ is the factorial of the multicomponent index $\vec{\nu}$.
\item
$\vec{z}^{\vec{\nu}} = \prod \limits_{i=1}^d z_i^{\nu_i}$ is the $\vec{\nu}$-th power of the
multicomponent variable $\vec{z}$.
\item 
$D_{\vec{x}}^{\vec{\nu}}= \frac{\partial^{\mid \vec{\nu} \mid}}{\partial x_1^{\nu_1} \dots \partial
x_d^{\nu_d}}$, $\mid \vec{\nu} \mid >0$ is the $\vec{\nu}$-th derivative with respect to the
multicomponent variable $\vec{x}$. We note that in this notation $D_{\vec{x}}^{\vec{0}}$ is the 
identity operator.
\item
$\mid \mid \vec{z} \mid \mid = \max  \mid z_i \mid$ is the maximum value
component of the multicomponent variable $\vec{z}$ in the interval $1 \le i \le d$.
\item
For the $d$-dimensional index $\vec{l}=(l_1,\dots,l_d)$ ($l_1,\dots,l_d$ are integers) we have
$\vec{l} \le \vec{\nu}$ when $l_i \le \nu_i, i=1,\dots,d$. Then we define
$$
{\vec{\nu}\choose \vec{l}}  = \prod \limits_{i=1}^d {\nu_i \choose l_i} = \frac{\vec{\nu} !}{\vec{l}! (\vec{\nu}-\vec{l})!}.
$$
\item 
Ordering of vector indexes. For two vector indexes $\vec{\mu}=(\mu_1,\dots,\mu_d)$ and $\vec{\nu}=
(\nu_1,\dots,\nu_d)$ we have $\vec{\mu} \prec \vec{\nu}$ when one of the following holds
\begin{enumerate}
\item   
$\mid \vec{\mu} \mid < \mid \vec{\nu} \mid$.
\item
$\mid \vec{\mu} \mid = \mid \vec{\nu} \mid$ and $\mu_1 < \nu_1$.
\item
$\mid \vec{\mu} \mid = \mid \vec{\nu} \mid$, $\mu_1 = \nu_1$, $\dots$ $\mu_k = \nu_k$ and 
$\mu_{k+1} < \nu_{k+1}$ for some $1 \le k < d$.
\end{enumerate}
\end{enumerate}
Below we use also the notation
\begin{equation}\label{m2}
h_{(\vec{\nu})} = D_{\vec{x}}^{\vec{\nu}}h; \ \ \ f_{(\vec{\lambda})} = D_{\vec{y}}^{\vec{\lambda}}f;
\ \ \ g_{(\vec{\mu})}^{(i)} = D_{\vec{x}}^{\vec{\mu}}g^{(i)}; \ \ \
\vec{g}_{(\vec{\mu})} =(g_{(\vec{\mu})}^{(1)},\dots,g_{(\vec{\mu})}^{(m)}). 
\end{equation}
The Faa di Bruno formula for the composite derivative of a function containing functions of many variables is \cite{cs_gen}
\begin{equation}\label{main}
h_{(\vec{\nu})} = \sum \limits_{1 \le \mid \vec{\lambda} \mid \le n} f_{(\vec{\lambda})}
\sum \limits_{s=1}^n \sum \limits_{p_s (\vec{\nu}, \vec{\lambda})} (\vec{\nu}!)
\prod \limits_{j=1}^s \frac{[ \vec{g}_{(\vec{l}_j)} ]^{\vec{k}_j}}{(\vec{k}_j!) [\vec{l}_j!]^{\mid \vec{k}_j \mid}}.
\end{equation}
In (\ref{main}) $n= \mid \vec{\nu} \mid$. In addition,
\begin{equation}\label{m3}
p_s(\vec{\nu},\vec{\lambda}) = \{ \vec{k}_1,\dots,\vec{k}_s; \vec{l}_1,\dots,\vec{l}_s \}, \ \ \
\mid \vec{k}_i \mid >0 .
\end{equation}
Moreover,
\begin{equation}\label{m4}
0 \prec \vec{l}_1 \dots \prec \vec{l}_s, \ \ \ \sum \limits_{i=1}^s \vec{k}_i = \vec{\lambda}, \ \ \
\sum \limits_{i=1}^s \mid \vec{k}_i\mid \vec{l}_i = \vec{\nu}.
\end{equation}
\par 
(\ref{main}) can be simplified by a change of the notation \cite{cs_gen}. We introduce
\begin{equation}\label{m6}
p(\vec{\nu}, \vec{\lambda}) = \{\vec{k}_1, \dots,\vec{k}_n; \vec{l}_1,\dots,\vec{l}_n \}, \ \ \
1 \le s \le n,
\end{equation}
and,
\begin{eqnarray}\label{m7}
\vec{k}_i=0;  \ \ \vec{l}_i=0, \ \ \ 1 \le i \le n-s,
 \mid \vec{k}_i \mid >0, \ \ n-s +1 \le i \le n.
\end{eqnarray}
Finally
$ 0 \prec \vec{l}_{n-s+1} \dots \prec \vec{l}_n$ are such that $\sum \limits_{i=1}^n \vec{k}_i = \vec{\lambda}_i$ and $\sum \limits_{i=1}^n \mid \vec{k}_i \mid \vec{l}_i = \vec{\nu}$. Then
(\ref{main}) can be written as
\begin{equation}\label{mainx}
h_{(\vec{\nu})} = \sum \limits_{1 \le \mid \vec{\lambda} \mid \le n} f_{(\vec{\lambda})}
\sum \limits_{p (\vec{\nu}, \vec{\lambda})} (\vec{\nu}!)
\prod \limits_{j=1}^n \frac{[ \vec{g}_{(\vec{l}_j)} ]^{\vec{k}_j}}{(\vec{k}_j!) [\vec{l}_j!]^{\mid \vec{k}_j \mid}}.
\end{equation} 
\par
We discuss below the specific case when the composite function $h$ is a function of two independent
variables $x_1$ and $x_2$. In addition we consider the case of composite function containing two functions 
of two independent variables.
In this case the composite function is a function of the functions $g^{(1)}(x_1,x_2)$ and 
$g^{(2)}(x_1,x_2)$. 
The Faa di Bruno formula for composite function containing two functions which are functions of
two variables is
\begin{eqnarray}\label{main4}
h_{(\vec{\nu})} = \frac{\partial^{\nu_1 + \nu_2} h}{\partial x_1^{\nu_1} \partial x_2^{\nu_2}} = 
\sum \limits_{1 \le (\lambda_1 + \lambda_2) \le \nu_1 + \nu_2} 
\frac{\partial^{\lambda_1  + \lambda_2}f}{\partial {g^{(1)}}^{\lambda_1} \partial {g^{(2)}}^{\lambda_2}}
\Bigg \{ \sum \limits_{s=1}^{\nu_1 + \nu_2}
\sum \limits_{p_{s}(\vec{\nu}, \vec{\lambda})}(\nu_1 ! \nu_2 !) \times \nonumber \\
 \prod \limits_{j=1}^{s} 
\Bigg [\frac{1}{(k_{j,1} !  k_{j,2} !)(l_{j,1}!+l_{j,2}!)^{k_{j,1}+  k_{j,2}}}
\prod \limits_{i=1}^2 \left( \frac{\partial^{l_{j,1}+l_{j,2}}}{\partial x_1^{l_{j,1}}  \partial x_2^{l_{j,2}}} g^{(i)} \right)^{k_{j,i}} \Bigg] \Bigg \}.
\end{eqnarray}
The version of (\ref{main4}) occurring from (\ref{mainx}) is 
\begin{eqnarray}\label{main4x}
h_{(\vec{\nu})} = \frac{\partial^{\nu_1 + \nu_2} h}{\partial x_1^{\nu_1} \partial x_2^{\nu_2}} = 
\sum \limits_{1 \le (\lambda_1 + \lambda_2) \le \nu_1 + \nu_2} 
\frac{\partial^{\lambda_1  + \lambda_2}f}{\partial {g^{(1)}}^{\lambda_1} \partial {g^{(2)}}^{\lambda_2}}
\Bigg \{
\sum \limits_{p(\vec{\nu}, \vec{\lambda})}(\nu_1 ! \nu_2 !) \times \nonumber \\
 \prod \limits_{j=1}^{n} 
\Bigg [\frac{1}{(k_{j,1} !  k_{j,2} !)(l_{j,1}!+l_{j,2}!)^{k_{j,1}+  k_{j,2}}}
\prod \limits_{i=1}^2 \left( \frac{\partial^{l_{j,1}+l_{j,2}}}{\partial x_1^{l_{j,1}}  \partial x_2^{l_{j,2}}} g^{(i)} \right)^{k_{j,i}} \Bigg] \Bigg \}.
\end{eqnarray}
\section{An example of use of composite functions in SEsM}
We are going to show how the methodology of SEsM works
in presence of composite functions. We will use a specific form of the 
composite function:  composite function of a function of 2 variables $h=f[g^{(1)}(x,t),g^{(2)}(x,t)]$
\begin{equation}\label{e4}
h = \alpha + \beta_1 g^{(1)} + \beta_2 g^{(2)} + \gamma_1 {g^{(1)}}^2 + \gamma_2 {g^{(2)}}^2
+ \gamma_3 g^{(1)} g^{(2)}.
\end{equation}
\par
The example is connected to the equation
\begin{equation}\label{exam9}
(1+h^2)\left( \frac{\partial^2 h}{\partial x^2} - \frac{\partial^2 h}{\partial t^2} \right) - 2h\left[ \left(\frac{\partial h}{\partial x} \right)^2 - \left(\frac{\partial h}{\partial t} \right)^2\right] =
h(1-h^2).
\end{equation}
We apply SEsM and skip Step. 1 (no transformation of the nonlinearity) as the
noninearity of the equation is polynomial one. 
We have $h=f[g^{(1)}(x,t),g^{(2)}(x,t)]$. The needed derivatives  are as follows
\begin{equation}\label{der1}
\frac{\partial h}{\partial x} = \frac{\partial f}{\partial g^{(1)}} \frac{\partial g^{(1)}}{\partial x} + \frac{\partial f}{\partial g^{(2)}} \frac{\partial g^{(2)}}{\partial x},
\end{equation}
\begin{equation}\label{der2}
\frac{\partial h}{\partial t} = \frac{\partial f}{\partial g^{(1)}} \frac{\partial g^{(1)}}{\partial t} + \frac{\partial f}{\partial g^{(2)}} \frac{\partial g^{(2)}}{\partial t}.
\end{equation}
\begin{eqnarray}\label{der3}
\frac{\partial^2 h}{\partial x^2} = \frac{\partial^2 f}{\partial {g^{(1)}}^2} \left( \frac{\partial g^{(1)}}{\partial x}\right)^2 + 2 \frac{\partial^2 f}{\partial g^{(1)} \partial g^{(2)}}  \frac{\partial g^{(1)}}{\partial x} \frac{\partial g^{(2)}}{\partial x} +  \frac{\partial f}{\partial g^{(1)}} \frac{\partial^2 g^{(1)}}{\partial x^2} + \nonumber \\
\frac{\partial^2 f}{\partial {g^{(2)}}^2} \left( \frac{\partial g^{(2)}}{\partial x}\right)^2 + 
 \frac{\partial f}{\partial g^{(2)}} \frac{\partial^2 g^{(2)}}{\partial x^2}.
\end{eqnarray}

\begin{eqnarray}\label{der4}
\frac{\partial^2 h}{\partial t^2} = \frac{\partial^2 f}{\partial {g^{(1)}}^2} \left( \frac{\partial g^{(1)}}{\partial t}\right)^2 + 2 \frac{\partial^2 f}{\partial g^{(1)} \partial g^{(2)}}  \frac{\partial g^{(1)}}{\partial t} \frac{\partial g^{(2)}}{\partial t} +  \frac{\partial f}{\partial g^{(1)}} \frac{\partial^2 g^{(1)}}{\partial t^2} + \nonumber \\
\frac{\partial^2 f}{\partial {g^{(2)}}^2} \left( \frac{\partial g^{(2)}}{\partial t}\right)^2 + 
 \frac{\partial f}{\partial g^{(2)}} \frac{\partial^2 g^{(2)}}{\partial t^2}.
\end{eqnarray}

(\ref{exam9}) becomes
\begin{eqnarray}\label{exam10}
(1+h^2) \Bigg \{ \frac{\partial^2 f}{\partial {g^{(1)}}^2} \Bigg[ \left( \frac{\partial g^{(1)}}{\partial x}\right)^2  - \left( \frac{\partial g^{(1)}}{\partial t}\right)^2\Bigg]
+ 2 \frac{\partial^2 f}{\partial g^{(1)} \partial g^{(2)}}  \Bigg[ \frac{\partial g^{(1)}}{\partial x} \frac{\partial g^{(2)}}{\partial x} - \frac{\partial g^{(1)}}{\partial t} \frac{\partial g^{(2)}}{\partial t} \Bigg] +  \nonumber \\
\frac{\partial f}{\partial g^{(1)}} \Bigg[ \frac{\partial^2 g^{(1)}}{\partial x^2}  - \frac{\partial^2 g^{(1)}}{\partial t^2}\Bigg]+ 
\frac{\partial^2 f}{\partial {g^{(2)}}^2} \Bigg[ \left( \frac{\partial g^{(2)}}{\partial x}\right)^2 -  \left( \frac{\partial g^{(2)}}{\partial t}\right)^2\Bigg] + \nonumber \\
 \frac{\partial f}{\partial g^{(2)}} \Bigg[ \frac{\partial^2 g^{(2)}}{\partial x^2} -  \frac{\partial^2 g^{(2)}}{\partial t^2} \Bigg]
  - 2h \Bigg \{  \left( \frac{\partial f}{\partial g^{(1)}}\right)^2 \Bigg[ \left( \frac{\partial g^{(1)}}{\partial x}\right)^2 -  \left( \frac{\partial g^{(1)}}{\partial t}\right)^2 \Bigg]+ \nonumber \\
 \left( \frac{\partial f}{\partial g^{(2)}}\right)^2 \Bigg[ \left( \frac{\partial g^{(2)}}{\partial x}\right)^2 - \left( \frac{\partial g^{(2)}}{\partial t}\right)^2  \Bigg] + 2 \frac{\partial f}{\partial g^{(1)}} \frac{\partial f}{\partial g^{(2)}} \Bigg[ \frac{\partial g^{(1)}}{\partial x}\frac{\partial g^{(2)}}{\partial x} -
 \frac{\partial g^{(1)}}{\partial t}\frac{\partial g^{(2)}}{\partial t} \Bigg]\Bigg \} = h(1-h^2)
\end{eqnarray}

\par
The composite function $h$
is of the kind (\ref{e4}) where $\alpha=0$, $\beta_1 = \beta_2 =1$, $\gamma_1 = \gamma_2 =0$. In addition $g^{(1)}$ does not depend on $t$ and $g^{(2)}$ does not depend on $x$. 
Let $\gamma_3 = A$. The composite function becomes
\begin{equation}\label{exam11}
h(x,t) = A g^{(1)}(\alpha x)g^{(2)}(\delta \gamma t), \ \ \ \delta = \pm 1.
\end{equation}
The composite function (\ref{exam11}) allows for complicated simple equations for $g^{(1)}$ and $g^{(2)}$.
These equations can be of the kind of equations for the elliptic functions of Jacobi:
\begin{eqnarray}\label{exam12}
\left(\frac{dg^{(1)}}{dx}\right)^2 &=& \alpha(a_1 {g^{(1)}}^4 + b_1 {g^{(1)}}^2 + c_1) \nonumber \\
\left(\frac{dg^{(2)}}{dx}\right)^2 &=& \beta(a_2 {g^{(2)}}^4 + b_2 {g^{(2)}}^2 + c_2).
\end{eqnarray}
Because of all above, (\ref{exam9}) is reduced to a system of algebraic equations
\begin{eqnarray}\label{exam13}
\alpha^2 b_1 -\gamma^2  b_2 &=& 1 \nonumber \\
\alpha^2 a_1 + \gamma^2 A^2 c_2 &=& 0 \nonumber \\
\gamma^2  a_2 + \alpha^2 A^2 c_1 &=& 0.
\end{eqnarray}
(\ref{exam13}) has various non-trivial solutions (Step 7 of SEsM).  For an example, one of these solutions is
when $\alpha^2 - \gamma^2 < 1$. We can consider $A$ as a free parameter. 
Then  $\alpha^2  = \gamma^2 + \frac{A^2-1}{A^2+1} $. Thus,
\begin{eqnarray}\label{exam14}
h (x,t) = A {\rm cn} \left\{\alpha x; \frac{A^2[\alpha^2(A^2+1)+1]}{\alpha^2(A^2+1)^2} \right\} {\rm cn} \left\{ \delta \gamma t; \frac{A^2[\gamma^2(A^2+1)-1]}{\gamma^2(A^2+1)^2} \right\} . 
\end{eqnarray}
In (\ref{exam14}) ${\rm cn} (\alpha x;k_1)$ and ${\rm cn} (\gamma t;k_2)$ are corresponding
Jacobi elliptic functions of modulus $0 \le k_1 \le 1$ and $0 \le k_2 \le 1$ respectively.
\par 
(\ref{exam13}) has an interesting specific case when $k_1=1$ and $k_1=0$. Then ${\rm cn}(\alpha x;k_1)=
{\rm sech}(\alpha x)$ and ${\rm cn}(\delta \gamma t) = \cos (\delta \gamma t)$. Then,
\begin{equation}\label{exam15}
h(x,t) = \frac{\cos \left[ \frac{\delta}{(A^2+1)^{1/2}} \right] t}{\cosh \left[\frac{A^2}{A^2+1} \right]^{1/2} x} .
\end{equation}
(\ref{exam15}) can be obtained also straightforward on the basis of the composite function
(\ref{exam11}) if one takes for $g^{(1)}$ and $g^{(2)}$ the corresponding simple equations for the
trigonometric and hyperbolic function respectively.
\par 
There are many other possible solutions. Several other examples are as follows. Let $k_1=k_2=1$ and $\alpha^2=1/(1-A)$; $\gamma^2 = A^2/(1-A^2)$.
Then,
\begin{equation}\label{exam16}
h(x,t) = a \frac{\sinh\left[(1/(1-A^2))^{1/2}x \right]}{\cosh \left[ (A^2/(1-A^2))^{1/2} t\right]}
\end{equation}
We note that this solution is specific case of the more complicated solution
\begin{equation}\label{exam17}
h(x,t) = A \frac{{\rm sn}(\alpha x;k_1)}{{\rm cn}(\alpha x;k_1)} {\rm dn}(\delta \gamma t; k_2),
\end{equation}
where
$$
k_1^2 = \frac{\alpha^2(1-A^2)^2+A^2}{\alpha^2(1-A^2)}; \ \ \
k_2^2 = \frac{A^2 - \gamma^2(1-A^2)^2}{\gamma^2A^2(1-A^2)}; \ \ \
\gamma^2 = \alpha^2 A^2
$$
Another example is when
\begin{equation}\label{exam18}
k_1^2=1 - \frac{1-\alpha^2(A^2+1)/A^2}{\alpha^2(A^2+1)}; \ \ \ 
k_2^2 = \frac{A^2[1-\gamma^2(\alpha^2+1)]}{\gamma^2(A^2+1)}; \ \ \
\alpha^2 = A^2 \gamma^2
\end{equation}
The corresponding solution is
\begin{equation}\label{exam19}
h(x,t)=A {\rm dn}(\alpha x, k_1) {\rm sn}(\delta \gamma t; k_2)
\end{equation}
(\ref{exam19}) has a specific case when $k_1=[1-1/A^4]^{1/2}$, and $k_2=1$.
In this case $\alpha^2 = A^2/(A^2+1)$, and $\gamma^2 = A^2/(A^2+1)$.
The solution becomes
\begin{equation}\label{exam20}
h(x,t)=A {\rm dn} \left[ \frac{A^2x}{A^2+1}; \left(1- \frac{1}{A^4} \right)^{1/2} \right] 
\tanh \left[\delta \left( \frac{At}{A^2+1}  \right) \right] 
\end{equation}
\par 
Another solution of the system of nonlinear algebraic equations is
\begin{equation}\label{exam21}
k_1^2 = 1 - \frac{\alpha^2(A^2-1)/A^2 -1)}{\alpha^2(A^2-1)^2}; \ \ \
k_2^2 = 1- \frac{A^2[\gamma^2(A^2-1)-1]}{\gamma^2(A^2-1)}; \ \ \
\alpha^2 =A^2 \gamma^2
\end{equation}
The corresponding solution is
\begin{equation}\label{exam22}
h(x,t)= A {\rm dn}(\alpha x; k_1) \frac{{\rm sn}(\delta \gamma t; k_2)}{{\rm cn}(\delta \gamma t; k_2)}
\end{equation}
Here we have again the specific solution $k_1=(1-1/A^4)^{1/2}$,$k_2=0$. Then $\alpha^2 = A^2/(A^2-1)$ and
$\gamma^2 = 1/(A^2-1)$. The solution is 
\begin{equation}\label{exam23}
h(x,t) = A {\rm dn} \left[\frac{A^2 x}{A^2-1}; \left(1 - \frac{1}{A^4} \right)^{1/2} \right]
\tan \left[ \left( \frac{A^2}{A^2-1} \right)^{1/2}t \right]
\end{equation}
\section{Concluding remarks}
Yhis paper is devoted to an aspect  of the application of the Simple Equations Method (SEsM) for obtaining excat soluions of nonliner differential equations.
This aspect is the use of composite function in the process of the application of the methodology. The need of use of composite functions in SEsM occurs in a natural way.
The reason is that the searched solution of the solved equation has to be constructed as a composite function of  functions which are solutions of more simple differential equations.  This leads to the need of use of the Faa di Bruno relationship for the derivatives of a composite functions. This use of composite
functions in the methodology of SEsM opens a possibility for obtaining additional results on the methodology 
as well as specific solutions of many nonlinear differential equations.


\begin{thebibliography}{99}
\bibitem{a1}
R. Axelrod, M. Cohen. \emph{Harnessing Complexity} (Basic Books, New York, 2001)
\bibitem{aa1}
P. G. Drazin. \emph{Nonlinear Systems} (Cambridge University Press, Cambridge, UK, 1992).
\bibitem{a3}
R. Lambiotte, M. Ausloos. Journal of Statistical Mechanics: Theory and Experiment P08026 (2007)
\bibitem{cs4}
M. Mitchell. Artificial Intelligence \textbf{170}, 1194 -- 1212 (2006).
\bibitem{ax2}
R. Kutner, M. Ausloos, D. Grech, T. Di Matteo, C. Schinckus, H. E. Stanley.
Physica A \textbf{516}, 240 - 253 (2019).
\bibitem{a3x}
N.K. Vitanov, F. H. Busse. ZAMP \textbf{48}, 310 -- 324 (1997).
\bibitem{cs2}
R. M. May, S. A. Levin, G. Sugihara. Nature \textbf{451},  893 -- 895 (2008).
\bibitem{sheard}
S. A. Sheard, A. Mostashari. Systems Engineering, \textbf{12}, 295 - 311 (2009).
\bibitem{ay1}
M. Bahrami, N. Chinichian, A. Hosseiny, G. Jafari, M.  Ausloos.  Physica A 
\textbf{540}, 123203 (2020).
\bibitem{v98a}
N.K. Vitanov. Physics Letters A, \textbf{248}, 338-346, (1998)
\bibitem{lenpert}
R. J. Lempert. PNAS USA, {\bf 99}, Suppl. 3, 7309 - 7313.
\bibitem{ax4}
N. K. Vitanov, M. Ausloos, G. Rotundo. Advances in Complex Systems \textbf{15}, Supp. 01, 1250049 (2012)
\bibitem{f1}
J. Foster. Cambridge Journal of Economics \textbf{29}, 873 - 892 (2005).
\bibitem{ax5}
N. K.Vitanov, K. N. Vitanov.  Mathematical Social Sciences, \textbf{80}, 108 -- 114 (2016).
\bibitem{camer1}
L. Cameron, D. Larsen-Freeman. Journal of Applied Linguistics, \textbf{17}, 226 - 239 (2007).
\bibitem{a4}
L. A. N. Amaral, A. Scala, M. Barthelemy, H. E. Stanley.   Proceedings of the 
National Academy of Sciences, \textbf{97}, 11149 -- 11152 (2000).








\bibitem{n1}
J. K. Hale. \emph{Oscillations in Nonlinear Systems} (Dover, New York, 1991).
\bibitem{n4}
A. S. Pikovsky, D. L. Shepelyansky. Phys. Rev. Lett. \textbf{100}, 094101  (2008).
\bibitem{nd1}
I. P. Jordanov. Comp. rend. Acad. Sci. Bulg, \textbf{61}, 307 -- 314 (2008).
\bibitem{n7}
Y. Niu,  S. Gong. Phys. Rev. A \textbf{73}, 053811 (2006).
\bibitem{t1}
H. Kantz,  T. Schreiber. \emph{Nonlinear Time Series Analysis} (Cambridge University Press, Cambridge, UK, 2004).
\bibitem{td1}
R. Struble. \emph{Nonlinear Differential Equations}. (Dover, New York, 2018).
\bibitem{t5}
P. J. Brockwell, R. A. Davis, M. V. Calder. \emph{Introduction to Time Series and Forecasting}. (Springer, New York, 2002).
\bibitem{v00}
N. K. Vitanov. Physica D {\bf 136},  322 -- 339 (2000) 
\bibitem{fuchs}
A. Fuchs. \emph{Nonlinear Dynamics in Complex Systems}. (Springer, Berlin, 2014)
\bibitem{golgst}
S. Goldstein. Phychological Inquiry, {\bf 8} , 125 - 128 (1997).
\bibitem{vdk06}
N. K. Vitanov, Z. Dimitrova, H. Kantz. Physics Letters A,  \textbf{346} 350-355 (2006)
\bibitem{n2}
K. Kawasaki, T. Ohta. Kink dynamics in one-dimensional nonlinear systems. 
Physica A, \textbf{116}, 573 -- 593 (1982).
\bibitem{td3}
N. K. Vitanov, K. N. Vitanov.  Physica A, \textbf{527}, 121174 (2019).
\bibitem{td4}
N. K. Vitanov, R. Borisov, K. N. Vitanov. Physica A, {\bf 581}, 126207 (2021).
\bibitem{t10}
H. Sedaghat. \emph{Nonlinear Difference Equations: Theory with Applications to Social Science Models}  (Springer Science \& Business Media, Dordrecht, 2013)





\bibitem{hopf}
E. Hopf. Communications on Pure and Applied Mathematics, \textbf{3},  201 -- 230 (1950).
\bibitem{cole}
J. D. Cole. Quarterly of Applied Mathematics \textbf{9},   225 -- 236 (1951).	
\bibitem{ablowitz}
M. J. Ablowitz, D. J. Kaup, A. C. Newell, H. Segur. Studies in Applied  Mathematics,  \textbf{53},  
249 -- 315 (1974) .	
\bibitem{ac}
M. J. Ablowitz,  P. A. Clarkson. \emph{Solitons, Nonlinear Evolution Equations and Inverse Scattering}. (Cambridge University Press, Cambridge, UK, 1991).
\bibitem{gardner}
C. S. Gardner,  J. M. Greene, M. D. Kruskal, R. R. Miura.  Phys. Rev. Lett. \textbf{19}, 
1095 -- 1097 (1967).
\bibitem{hirota}
R. Hirota.   Phys. Rev. Lett. \textbf{27},   1192 -- 1194 (1971).
\bibitem{hirota1}
R. Hirota. \emph{The Direct Method in Soliton Theory}. (Cambridge University Press, Cambridge, UK, 2004).
\bibitem{v23}
N. K. Vitanov.  {\em Axioms}, {\bf 12}, No. 12, 1106 (2023).



\bibitem{v20}
N. K. Vitanov, Z. I. Dimitrova, K. N. Vitanov.  {\em Entropy}, {\bf 23}, No.1, 10 (2020).
\bibitem{v22}
N. K. Vitanov,   {\em Entropy}, {\bf 24}, No. 11, 1653,(2022).
\bibitem{v22a}
N. K. Vitanov.{\em  AIP Conference Proceedings} {\bf 2459}, 020003  (2022).




\bibitem{k05}
N. A. Kudryshov. Chaos, Solitons \& Fractals \textbf{24},   1217 -- 1231 (2005).
\bibitem{kl08}
N. A. Kudryashov, N. B. Loguinova.  Applied Mathematics and Computation \textbf{205}, 361 -- 365 (2008).
\bibitem{se1}
N. K. Vitanov. Pliska Studia Mathematica Bulgarica, \textbf{30}, 29 -- 42 (2019).
\bibitem{se2}
N. K. Vitanov. Journal of Theoretical and Applied Mechanics, \textbf{49}, 107 -- 122 (2019).
\bibitem{se3}
N. K. Vitanov.  AIP Conference Proceedings \textbf{2159},  030038 (2019). 
\bibitem{se4}
N. K. Vitanov, Z. I.  Dimitrova.  AIP Conference Proceedings \textbf{2159},  030039 (2019). 






\bibitem{mv1}
N. Martinov, N. Vitanov. Journal of Physics A: Mathematical and General \textbf{25},  L51 -- L56 (1992).
\bibitem{mv21}
N. Martinov, N. Vitanov. J. Phys A: Math. Gen. \textbf{25}, 3609 -- 3613 (1992)
\bibitem{mv4}
N. K. Martinov, N. K. Vitanov. Canadian Journal of Physics, \textbf{72}, 618 -- 624 (1994).
\bibitem{mv42}
N. Martinov, N. Vitanov. Zeitschrift f{\"u}r Physik B. \textbf{100}, 
129 -- 135 (1996).
\bibitem{mv5}
N. K. Vitanov. Journal of Physics A: Mathematical and General, \textbf{29},  5195 -- 5207 (1996).







\bibitem{1}
N. K. Vitanov, I. P. Joranov, Z. I. Dimitrova.  Communications in Nonlinear Science and Numerical Simulation 
\textbf{14}, 2379 -- 2388 (2009).
\bibitem{2}
N. K. Vitanov, I. P. Jordanov, Z. I. Dimitrova.   Applied Mathematics and Computation \textbf{215}, 2950-- 2964 (2009).
\bibitem{v10}
N. K. Vitanov. Communications in Nonlinear Science and Numerical Simulation
{\bf 15}, 2050 -- 2060 (2010).
\bibitem{vd10}
N. K. Vitanov,  Z. I. Dimitrova. Communications in Nonlinear Science and Numerical Simulation {\bf 15}, 2836 -- 2845  (2010).
\bibitem{vdk}
N. K. Vitanov, Z. I. Dimitrova, H. Kantz. Applied Mathematics and Computation, \textbf{216},  
2587 -- 2595 (2010).
\bibitem{v11}
N. K. Vitanov. Communications in Nonlinear Science and Numerical Simulation, \textbf{16}, 1176 -- 1185 (2011).


\bibitem{v11a}
N. K. Vitanov, Z. I. Dimitrova, K. N. Vitanov. Communications in Nonlinear Science and Numerical Simulation, \textsc{16},  3033 -- 3044 (2011).
\bibitem{v11b}
N. K. Vitanov.  Communications in Nonlinear Science and Numerical Simulation, \textsc{16},  4215 -- 4231 (2011).
\bibitem{pliska1}
N. K. Vitanov. Pliska Studia Mathematica Bulgarica \textbf{21}, 257 -- 266 (2012).

\bibitem{vd14}
N. K. Vitanov, Z. I. Dimitrova. Applied Mathematics and Computation, \textbf{247},    213 -- 217 (2014).
\bibitem{vdv17}
N. K. Vitanov, Z. I. Dimitrova, T. I. Ivanova.  Applied Mathematics and Computation, \textbf{315},  372 -- 380 (2017).


\bibitem{vdv15}
N. K. Vitanov, Z. I. Dimitrova, K. N. Vitanov.  Applied Mathematics and Computation, \textbf{269},  363 -- 378 (2015).
\bibitem{vd18}
N. K. Vitanov,Z. I. Dimitrova. Journal of Theoretical and Applied Mechanics, Sofia, \textbf{48}, No. 1, 59  -- 68 (2018).



\bibitem{n17}
N. K. Vitanov.  AIP Conference Porceedings, \textbf{2321}, 030035 (2021).
\bibitem{vnew21x}
N. K. Vitanov, Z. I. Dimitrova, K. N. Vitanov. Entropy, \textbf{23}, 10 (2021).
\bibitem{cs_gen}
G. M. Constantine, T. H. Savits.  Transactions 
of the American Mathematical Society, {\bf 348}, 503 -- 520 (1996).









\end{thebibliography}
\end{document}